# Planetary Exploration Using Cubesat Deployed Sailplanes

Adrien Bouskela[1], Alexandre Kling[3], Aman Chandra[2], Tristan Schuler[2],
Sergey Shkarayev[1], Jekan Thangavelautham[2]

[1]*Micro Air Vehicles Laboratory, Department of Aerospace and Mechanical Engineering, University of Arizona, 1130 N Mountain Ave, Tucson, Arizona, United States, 85721*

[2]*Space and Terrestrial Robotic Exploration Laboratory, University of Arizona, Department of Aerospace and Mechanical Engineering, 1130 N Mountain Ave, Tucson, Arizona, United States, 85721*

[3]*NASA Ames Research Center*

**Abstract**

Exploration of terrestrial planets such as Mars are conducted using orbiters, landers and rovers. Cameras and instruments onboard orbiters have enabled global mapping of Mars at low spatial resolution. Landers and rovers such as the Mars Science Laboratory (MSL) carry state-of-the-art instruments to characterize small localized areas. This leaves a critical gap in exploration capabilities: mapping regions in the hundreds of kilometers range.

A high science return/low cost solution is to deploy one or more sailplanes in the Martian atmosphere as secondary payloads deployed during Entry, Descent and Landing (EDL) of a MSL-class vehicle. These are packaged into 12U/24kg CubeSats, occupying some of the 190 kg of available ballasts. Sailplanes extend inflatable-wings to soar without power limitations by exploiting atmospheric features such as thermal updrafts for static soaring, and wind gradients for dynamic soaring. Such flight patterns have been proven effective on Earth, and demonstrated similarities between Earth and Mars show strong potential for a long lasting airborne science platform on Mars. The maneuverability of sailplanes offer distinct advantages over other exploration vehicles: they provide continuous reconnaissance of areas of interest from multiple viewpoints and altitudes with dedicated science instruments, achieving higher pixel-scale resolutions than orbital assets and enabling exploration capabilities over rugged terrain such as Valles Marineris, steep crater walls and the Martian highlands that remain inaccessible for the foreseeable future due to current EDL technology limitations.

In this paper, we extend our work on CubeSat-sized sailplanes with detailed design studies of different aircraft configurations and payloads, identifying generalized design principles for autonomous sailplane-based surface reconnaissance and science applications. We further analyze potential wing deployment technologies, including conventional inflatables with hardened membranes, use of composite inflatables, and quick-setting foam. We perform detailed modeling of the Martian atmosphere and possible flight patterns at Jerezo crater using the Mars Regional Atmospheric Modeling System (MRAMS) to provide realistic atmospheric conditions at the landing site for NASA's 2020 rover. We revisit the feasibility of the Mars Sailplane concept, comparing it to previously proposed solutions, and identifying pathways to build laboratory prototypes for high-altitude Earth based testing. Finally, our work will analyze the implications of this technology for exploring other planetary bodies with atmospheres, including Venus and Titan.

## 1. Introduction

A large number of sites of interests on Mars remain inaccessible due to current technological limitations in precision Entry, Descent and Landing (EDL) and inability to land in high-altitude and rugged terrains. Previously, several airborne missions providing access to these remote sites have been proposed for Mars [1-5]. These previous concepts are all powered and the inclusion of electric or chemical propulsion comes with great penalty on the mass and complexity of the flying vehicle. To address the challenge of long duration flight (few hours to a few days) with limited resources, nature provides an elegant source of inspiration in the form of the albatross, a bird able to sustain flight for thousands of miles over the ocean without ever making land. The secret of albatross's flight lies in utilizing its environment through dynamic soaring maneuvers. We conducted numerical simulations that showed that this maneuver is possible on Mars (Fig. 14).

Our primary concept features an orbiter with a launch mass typical of flagship Mars Science Laboratory (MSL)-class vehicle, which is placed into a high inclination Mars orbit. The orbiter carries several entry capsules with individual heatshields, parachutes, and deploys 8 sailplanes at different times and locations, including a reference science target in Melas Chasma, Valles Mariners, and a high latitude target in the summer hemisphere. Each sailplane is equipped with an MSSS ECAM-C50 reconnaissance camera [6], NASA JPL IRIS v2 X-band communications, a flight control system, and environmental instruments, totaling 5 kg. At 7 km altitude, the sailplanes are released and inflate their wings. The sailplanes fly with no power limitations by exploiting atmospheric wind gradients for dynamic soaring as well as thermal and slope flows for static soaring. High-maneuverability and low-altitude passes of sailplanes provide reconnaissance images of targets from multiple viewpoints. Sequential and simultaneous deployments allows in-situ monitoring of the winds, trace gasses, and investigation of temporal and spatial variabilities. At the end of the primary reconnaissance missions, the sailplanes are soft landed and continue collecting environmental data at the landing site.

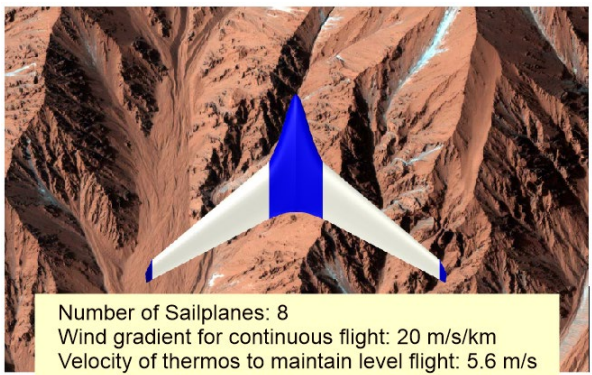

Fig. 1. Sailplane for Science Reconnaissance on Mars

The idea of a heavier-than-air craft for Mars exploration holds much promise for future discoveries on the planet. A small helicopter designed by NASA [7] is scheduled to launch onboard Mars 2020 in July 2020 and will attempt to perform first flight on another planet. The helicopter has two counter-rotating rotors and has a total mass of 1.8 kg. It is expected to fly up to five flights of only 1.5 min [8].

The autonomous aircraft ARES (Aerial Regional-scale Environmental Survey) was proposed by NASA [9]. Several propulsion technologies were considered for powering this vehicle including electrical motors, internal combustion and rocket systems. The aircraft intended to fly about an hour performing atmospheric probe missions.

An unconventional craft called Entomopter was designed in [10]. It uses two sets of flapping wings mounted on the fuselage. The wings generate insect-like motions that produce a lift and thrust force. The flight endurance is predicted to be 10-15 min.

Note that the flight endurance of the aircrafts proposed in the previous studies has been limited by the amount of fuel carried on board. Alternatively, the unpowered glider Prandtl-m [11] was designed based on the high-aspect-ratio flying wing. It is foldable and fits into a CubeSat. However, the glider is capable of only about 10 minutes of flight.

Among limitations and challenges of the previous research and development efforts, the short flight endurance is the most critical one. To overcome this underperformance, we propose to design an unpowered sailplane, employing dynamic and static soaring methods for flight in the Martian atmosphere.

Dynamic soaring of sailplanes in the Earth's atmospheric boundary layer has been well studied by using simulations and flight tests. Analysis of work-energy relationship of a sailplane in the wind is presented in [12]. The study shows that the energy neutral cycle depends on the maximum lift-to-drag ratio of the vehicle and the wind speed gradient. For a continuous wind profile, the minimum gradient of the wind was calculated providing the neutral energy cycle.

The dynamic soaring maneuvers of unmanned aerial vehicles in the atmospheric boundary layer were examined numerically for a number of different wind profiles [13]. Both 3-DOF and 6-DOF models were employed [14] to demonstrate dynamic soaring with extreme climbs to high altitudes in high wind conditions. The flight demonstrations on Earth [15, 16] proves its effectiveness in achieving long-endurance flights.

In our previous work, [17], we conducted a preliminary feasibility study of sailplane deployment and dynamic soaring in the Martian atmosphere. Finding that the lower gravity and atmospheric pressure did not hinder an aircraft's ability to perform the maneuver, and that higher flight speed, consequentially higher distances, are necessary due to the low Reynolds number conditions.

The current work continues the study of long endurance flight sustained by atmospheric conditions, including static soaring, and expands on the sailplane's design and the missions it can conduct. Continuous wind models are extracted from simulations conducted with the Mars Regional Atmospheric Modeling System (MRAMS) thus enabling advancement in solution fidelity. Several different deployment strategies are explored including quick deployment during EDL and deployment using a carrier hot-air balloon or blimp.

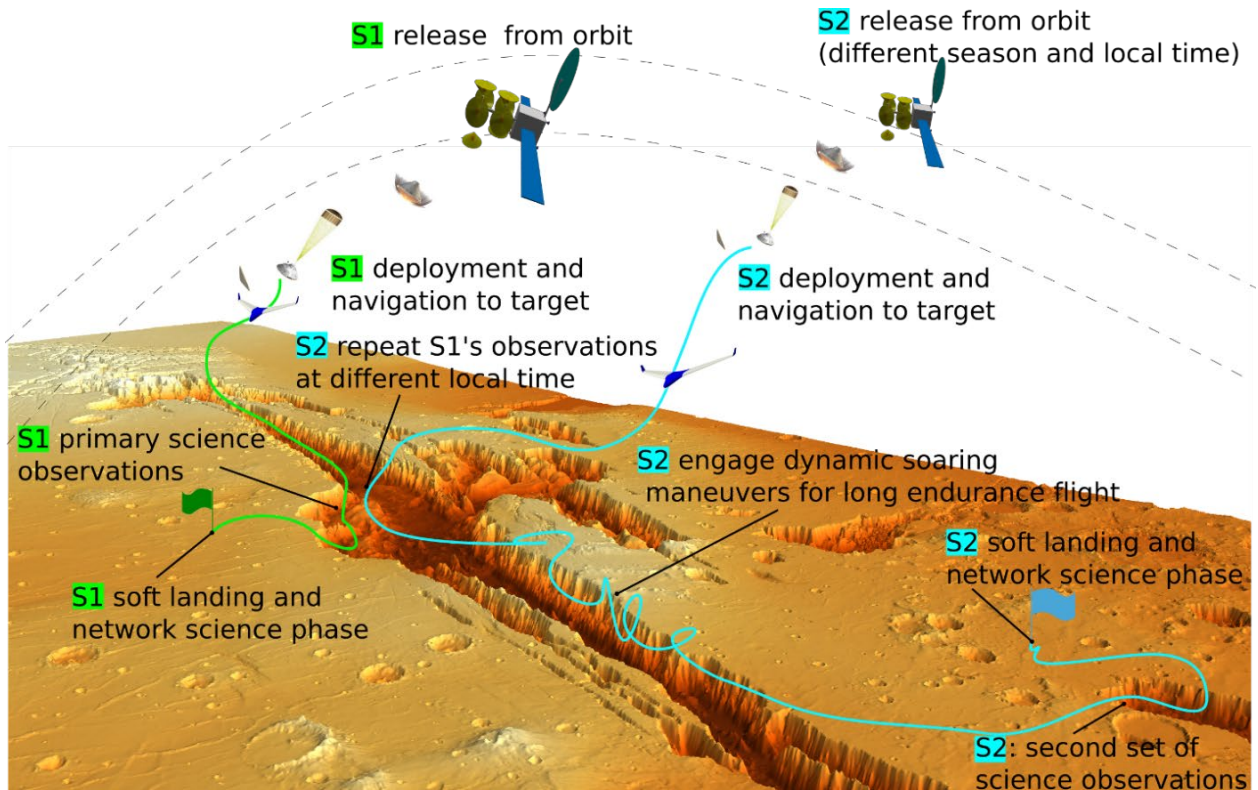

Fig. 2. Concepts of Operations for two sailplanes S1 and S2 deployed sequentially.

## 2. Airborne Mission

The mission can be broken down into five successive phases: 1) the initial deployment and leveling, 2) direct navigation to science target, possibly using some large scale currents, 3) search, location and observation of a science target (possibly multiple passes), 4) station keeping and long distance navigation using soaring maneuvers to secondary targets and finally 5) landing and extended surface meteorology operations. Several sailplanes can be deployed simultaneously.

Figure 2 shows the notional concepts of operations for two sailplanes S1 and S2, released at different local times and seasons. S1 follows a simple mission profile with a direct glide to image a science target in Melas Chasma, followed by a soft landing on the Southern highlands. S2 follows a more complex profile, repeating S1 observations at a different local time, engaging in thermal and dynamic soaring maneuvers to cover long distances to reach a second science target, and soft-lands at a different location. Both sailplanes monitor the winds, turbulence, presence of trace gases and perform a joint, network-based meteorology experiment at the end of their primary reconnaissance mission.

Figure 3 describes the mission concept at the Valles Marineris targets. S3 would deploy and navigate to the canyon floor, perform sustained flight and science imaging with ground proximity dynamic soaring, either advancing down the canyon or loitering over local targets of interest.

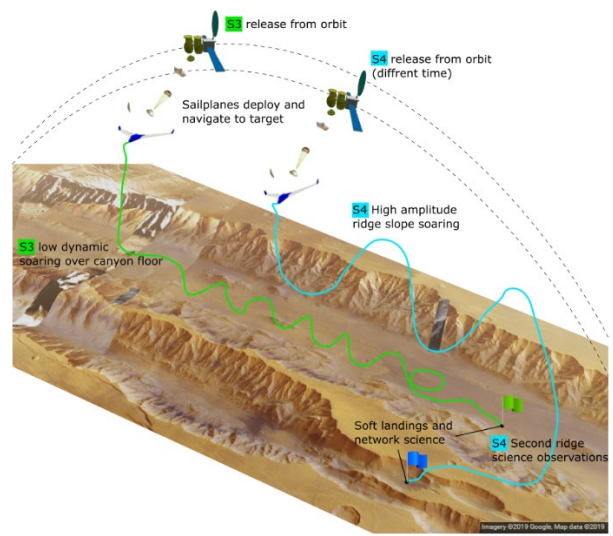

Fig. 3. Concept of Operations from S3 and S4 sailplanes.

Images and wind measurements collected in previous flights will provide insight for S4, which would deploy at a later local time to fly high amplitude slope soaring trajectories along the canyon wall, surveying the cliff sides with multiple climbing and descending passes, finishing with enough energy for a secondary target. Gliding along the wall face for the entire length of the canyon is also possible in the case that the local conditions are unfavorable to energy conservation or increase. Each primary mission terminates with a soft landing, joining S1 and S2 in networking meteorological measurements.

### 3. Aerodynamic Design of Sailplane

The Martian atmosphere is hundred times thinner than that on Earth ($\rho$ = 0.0137 kg/m$^3$). Even though the gravity on the Mars is significantly lower, $g$ = 3.7278 kg m/s$^2$, it still makes it an endeavor to achieve sustainable flight on this planet. With a 'sea' level viscosity of $\mu$ = 1.08·10$^{-5}$ N s/m$^2$, a craft with an assumed wing chord of 0.54 m and a speed of 70 m/s, the Reynolds number is 47,950. This value falls within the low Reynolds number range (low Re is less than 2·10$^5$). In contrast, the Mach number in flight can reach 0.8. This case has not been extensively researched, with existing studies for a limited number of airfoils and airframes [18,19]. To mitigate these uncertainties for the presented design, future wind tunnel experiments will be conducted, investigating the aerodynamics of airfoils at relatively low Reynolds and moderate Mach numbers pertinent to Mars airborne missions.

Design problems associated with low atmospheric density and low Reynolds number are considered, and a number of thin, cambered airfoils [20] were analyzed and the airfoil S9033 was selected due to its high lift-to-drag ratio and moderate, positive pitching moment coefficient. A sweptback tailless design provides better performance in comparison to a conventional fixed-wing aircraft with a tail. The proposed sailplane design is shown in Fig. 4. It features a blended-wing-body. A notable feature of this configuration is the lift-generating fuselage, which is seamlessly blended with the body. Not only is the lift produced by the entire craft increased, the interference drag is reduced. To determine the best parameters of this design, an optimization of the aerodynamic efficiency was undertaken in our previous study [21]. The specifications of the sailplane are given in Table 1.

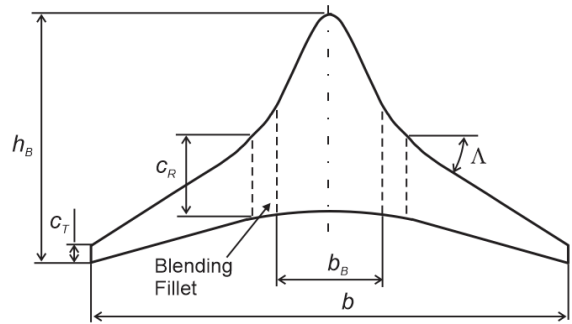

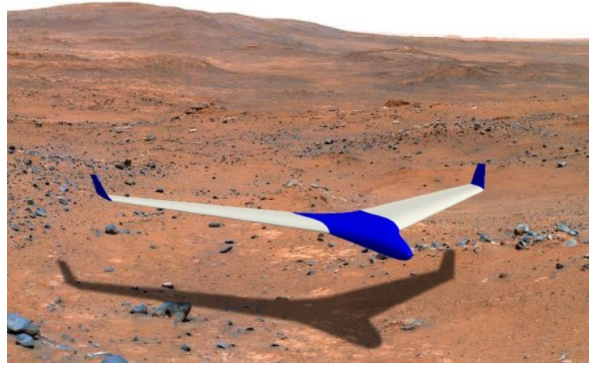

Fig. 4. Sailplane Planform (top) and 3D-rendering (bottom).

Table 1. Specifications of Mars sailplane.

| Parameter | Value | Parameter | Value |
|---|---|---|---|
| Wing/Body Airfoil | S9033 | Wing Area, m$^2$ | 1.8 |
| $b$, m | 3.35 | $C_T$, m | 0.34 |
| $b_B$, m | 0.7 | $h_B$, m | 1.12 |
| $C_R$, m | 0.56 | $\alpha$, deg | 45 |
| Aspect Ratio | 10 | Total Mass, kg | 5.0 |
| Flight Endurance | 21 min (min) to days (max) | Average Speed, m/s | 70 |
| | | Min. Sink speed, m/s | 6.28 |

### 4. Flight Simulations in Martial Atmosphere

Specific flights will be planned according to mission requirements. They will include the following patterns: dynamic soaring, gliding, soaring in updrafts, and dive and climbing flights. Numerical simulations will demonstrate feasibility of these flight patterns under Mars atmospheric conditions.

Consider a flight path of a sailplane modeled as a point-mass, $m$, with three degrees of freedom. Fig. 5.

illustrates conventions for forces, angles, and velocities. There are three applied forces: lift, $L$, drag, $D$, and gravitational force, $mg$. The aerodynamic lift and drag are conventionally written as

$$L = C_L 0.5\rho V_a^2 S; \quad D = C_D 0.5\rho V_a^2 S \qquad (1)$$

The drag polar for the flying wing calculated by XFLR5, using the S9033 airfoil, flying wing geometry, and $C_L \in [C_{L_{min}}, C_{L_{max}}]$, bounded by the convergence range.

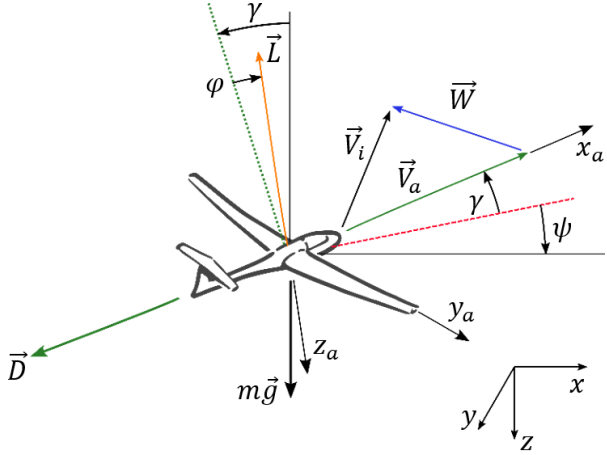

Fig. 5. Conventions for forces, angles, and velocities.

The dynamic soaring problem is presented within the inertial reference frame $(x, y, z)$, and the aerodynamic frame $(x_a, y_a, z_a)$ attached to the sailplane. The transformation from the inertial to the aerodynamic reference frame is performed using three canonical rotations about the yaw angle, $\psi$, the pitch, $\gamma$, and the roll, $\varphi$, Euler angles. The components of the velocity vector of the vehicle relative to atmosphere are:
$\vec{V}_a = V_a[\cos(\psi)\cos(\gamma), \cos(\gamma)\sin(\psi), -\sin(\gamma)]^T$ and components of wind in the inertial frame are denoted as $\vec{W} = [W_x\ W_y\ W_z]^T$. Then corresponding kinematical equations are

$$\vec{V}_i = \vec{V}_a + \vec{W} = \begin{bmatrix}\dot{x}\\\dot{y}\\\dot{z}\end{bmatrix} = V_a\begin{bmatrix}\cos(\psi)\cos(\gamma)\\\cos(\gamma)\sin(\psi)\\-\sin(\gamma)\end{bmatrix} + \begin{bmatrix}W_x\\W_y\\W_z\end{bmatrix} \qquad (2)$$

Equations of motion of the sailplane are obtained by applying the Newton's second law

$m(\dot{V}_a + \dot{W}_x \cos(\psi)\cos(\gamma) + \dot{W}_y \cos(\gamma)\sin(\psi) - \dot{W}_z \sin(\gamma)) = -D + mg\sin(\gamma)$

$m(\dot{\gamma} V_a - \dot{W}_x \cos(\psi)\sin(\gamma) - \dot{W}_y \sin(\psi)\sin(\gamma) - \dot{W}_z) = L\cos(\varphi) + mg\cos(\gamma)$

$m(\dot{\psi} V_a \cos(\gamma) - \dot{W}_x \sin(\psi) + \dot{W}_y \cos(\psi)) = L\sin(\varphi) \qquad (3)$

Herein, the inertial wind rates are:

$$\begin{aligned}\dot{W}_x &= \frac{\partial W_x}{\partial t} + \frac{\partial W_x}{\partial x}\dot{x} + \frac{\partial W_x}{\partial y}\dot{y} + \frac{\partial W_x}{\partial z}\dot{z}\\ \dot{W}_y &= \frac{\partial W_y}{\partial t} + \frac{\partial W_y}{\partial x}\dot{x} + \frac{\partial W_y}{\partial y}\dot{y} + \frac{\partial W_y}{\partial z}\dot{z}\\ \dot{W}_z &= \frac{\partial W_z}{\partial t} + \frac{\partial W_z}{\partial x}\dot{x} + \frac{\partial W_z}{\partial y}\dot{y} + \frac{\partial W_z}{\partial z}\dot{z}\end{aligned} \qquad (4)$$

Combining (2) and (3), the governing system of six first order differential equations can be written in the vector form:

$$\dot{\vec{Y}} = f(\vec{Y}, \vec{u}) \qquad (5)$$

where $\vec{Y} = [x, y, z, V_a, \gamma, \psi]^T$ and $\vec{u} = [C_L, \varphi]$ is the state vector and the control vector, respectively. The time dependent control parameters $C_L$ and $\varphi$ affect the flight trajectory by changing either lift magnitude or heading. It is sufficient to control climb, descend, and turns of the three degree of freedom model of the sailplane.

Performance of the trajectories is characterized with total energy $E = \Pi + T$ as the sum of potential energy $\Pi = -mgz$, and kinetic energy $T = 0.5mV_i^2$.

There are various types of flight needed for the planetary exploration purposes at selected sites. Based on the developed dynamic model and specific mission requirements, sailplane flight patterns can be devised and optimized. They comprise elements of gliding, static and dynamic soaring.

## 5. Gliding

Gliding is the simplest type of sailplane flight undertaken for reconnaissance missions. The main parameter characterizing this flight technique is the minimum sink rate. It shows the ability of the sailplane to perform long-range flights without assistance from atmospheric winds. For the current sailplane design, the minimum sink rate is found to be 6.83 m/s determined at the lift to drag ratio $C_L/C_D = 11.131$, $\alpha = 3.56°$ and $V_a = 76.4$ m/s. From the altitude of 6 km, the craft can glide 66.8 km one way. In the Valles Marineris, which depth is about 7 km, the distance travelled before soft landing would be 77.9 km or 17 minutes of flight time. For the reportedly deeper Melas Chasma these values are 100km and 21 minutes.

Gliding is the only possible flight method possible after direct deployment, where sailplanes cruise to sites with strong winds and harvest energy from the atmospheric environment.

## 6. Static soaring

Static soaring depends on the vertical winds and include thermal and slope soaring. Thermal soaring becomes available in the presence of temperature gradients, regardless of local terrain. They translate into vertical winds referred to as thermal updrafts and are the result of uneven heating or cooling of the large scale environment. They are known to exist on Mars thanks to measurements collected during spacecraft entries [22], the observation of winds and dust devils [23], as well as atmospheric simulations [24,25].

The measurements and simulations of ground winds [26 – 28] also infer that slope soaring is possible with prevailing winds accelerated and diverted around geological features such as cliffs, steep slopes and ridges. Causing locally vertical wind components, referred to as areas of lift or updrafts, and terrain dependent recirculation areas (Fig. 6.). These last ones are generally undesirable for soaring but under some conditions, a favorable shear layer can exist prior to a recirculation vortex, and is exploited with the later discussed dynamic soaring.

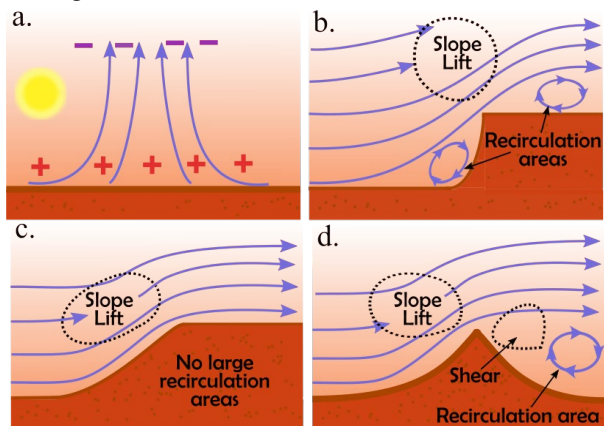

Fig. 6. Atmospheric flows for static soaring: (a) thermal updraft due to ground heating, (b) cliff, (c) steep slope, (d) and ridge subject to horizontal wind.

A sailplane can sustain flight if exposed to an updraft equal to its minimum sink velocity (6.28 m/s) and will gain energy if it's greater. Practically, updrafts are exploited until the weaker dissipating top part is reached, the sailplane then glides looking for another updraft using a predetermined search algorithm. Such patterns can help reach distant science targets by alternating thermal soaring, slope soaring, and gliding. Numerical modeling using the Large Eddy Simulation (LES) method, and thermal plume models adapted for Mars, and verified with recent ground based temperature measurements was conducted previously [29]. Finding that thermal updraft and downdraft structures represent a more significant portion of the dynamics in the Planetary Boundary layer (PBL). Results show pockets of 3 to 10 m/s of vertical velocity in the 1 to 6 km altitude range for the selected time, providing upwards of 40 km of gliding range to the sailplane, allowing for repeated harvesting while flying across an area. Further results estimate the presence of such dynamics over the entire equatorial region of the planet, supporting hypnotizes of possible long flight endurance using thermal features. To fully exploit this, guidance laws will need to be developed to avoid downdrafts which were also shown to be significant [29].

## 7. Flying canyons and ridges

There are several flight patterns presented in the previous sections: gliding, diving, pullout of dive, climbing and slope soaring. They provide close lateral viewing capabilities necessary for geological studies and the search for trace gases such as methane over ridges and canyon walls, such as Valles Marineris. They demonstrate that Mars sailplanes are suitable for close-to-wall-flying passes over locations inaccessible with traditional EDL architectures.

Diving patterns can vary in altitude (Fig. 7) with a steep dive (1a), pull-up maneuver at maximum load factor (1b), climb (1c), and slope soaring for energy gain (1d). Similarly, the updrafts caused can also sustain constant altitude flight or loitering in the lift region (Fig. 5).

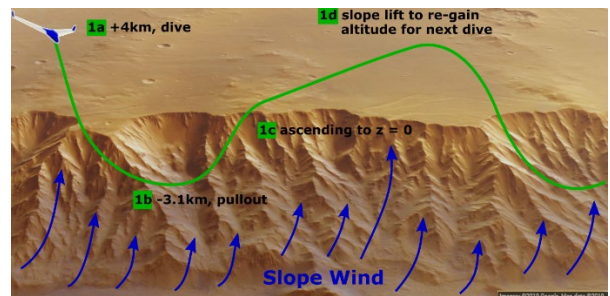

Fig. 7. Canyon exploration concept using slope soaring.

The trajectory is split into a dive and a climb section for numerical solving with the sailplane's dynamic model, both comprised of $\gamma = 0$ ($\dot{z} = 0$) initial and final conditions, where the final state and controls of the dive define the initial conditions of the climb. The system of equations can be simplified for a two-dimensional trajectory with $\psi = 0$ and no $\varphi$ control. The states are then $\vec{Y}' = [x, z, V_a, \gamma]^T$ with the differential problem $\vec{Y}' = f(\vec{Y}', \alpha)$, where $\alpha$ is the control.

In a dive, potential energy is converted into kinetic energy, with the sailplane's terminal velocity greater than $M=1$ a design speed limit is set to $M=0.8$ to prevent supersonic flight, translating to system boundaries:

$$V_{a\ stall} \leq V_a \leq V_{a\ design\ limit} \quad (6)$$

When the sailplane is in a dive (60 and 80 degrees pitch limit in Fig. 9a) there exists a maximum dive speed, referring to the maximum achievable velocity at the end of the dive, and is calculated to be 187.1 m/s through dive numerical simulations and the objective is the following:

$$-\gamma_{limit} \leq \gamma \leq 0; \quad x \geq 0; \quad z_f \geq z \geq z_i \quad (7)$$
$$\max_{C_L}\{V_a: t_f\} \quad (8)$$

The maximum dive speed sets an upper limit on the available kinetic energy to climb back up to $z=0$, hence a maximum achievable altitude and equal depth the sailplane can reach and return to the canyon ceiling. Numerical solutions for 'climb' (Fig. 8b) with boundaries and objective as stated below, show that with no vertical winds the maximum depth reachable is $-3138m\ (z = 3138m)$ with zero being the canyon ceiling, or $E = 2.881e5\ J$ as the minimum total energy needed to re-exit the canyon:

$$0 \leq \gamma \leq 90; \quad ; \quad x \geq 0; \quad z \geq z_i \quad (9)$$
$$\max_{C_L}\{z: t_f\} \quad (10)$$

To reach lower depths and initial energies, vertical winds must positively affect the sailplane to compensate for the difference in initial energy. In high wind conditions, lift can exist below zero, but commonly energy is gained through slope soaring above the canyon ceiling (Fig. 8). The minimum sink speed for static soaring is a function of $V_a$, and the calculated relationship sets a lower boundary on $V_{a\ target} = 55$ m/s in climb. Note that the sink velocity has a minimum at $V_a = 62.65$ m/s (Fig. 9), making it more efficient to first gain kinetic energy from the slope. Then, the added energy can be used to gain further potential energy towards the end of the climb where slope benefits are smaller, satisfying dive initial conditions.

Another option is diving at a larger initial velocity, allowing for less total traveled altitude since the dive maximum speed is reached earlier under those conditions (Fig. 8c).

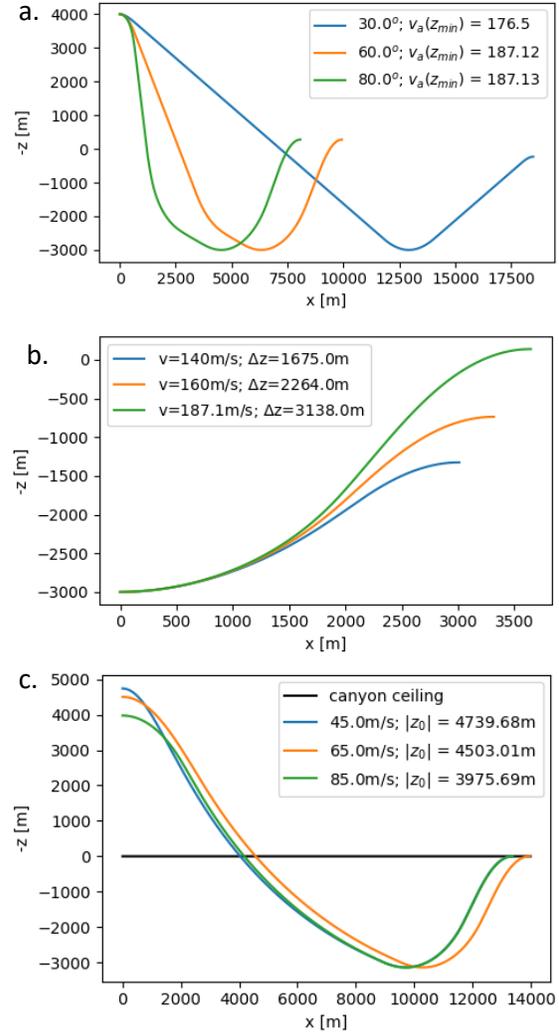

Fig. 8. Sailplane trajectories. (a) Complete dive, pull-up, and climb cycles with different lower pitch boundaries, (b) climb with different initial velocities, (c) and complete cycle with different initial velocities.

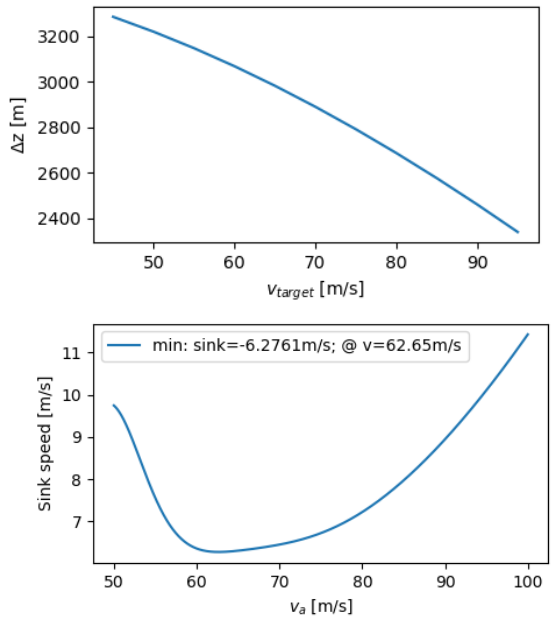

Fig. 9. Climb performance indicators, with the vertical distance vs. $V_{a\,target}$ (top), and the sink speed at $V_{a\,target}$ (bottom)

The sink velocity and trajectory results show that altitude oscillating slope soaring is possible in the Martian atmosphere around large features such as the Valles Marineris walls. Its reported depth is ~7000 m and current results show the sailplane flying past only about half its depth. Future research will attempt to characterize atmospheric flows that can be exploited to reach deeper trajectories, including the added kinetic energy from horizontal winds, studied in the context of dynamic soaring.

### 8. Dynamic soaring flight paths

Dynamic soaring flight paths are more complex, comprise typical cycles of climbing into a non-uniform wind, high-g turns and descent.

The sailplane will fly into a horizontal wind where vertical variations in magnitude are present, and by doing so, it will harvest additional kinetic and potential energy. This high-risk, high-reward method commonly used in nature by albatross birds [30] for long distance traveling and aero-modeling [31], has the potential to provide days of continues flight using advanced autonomous control.

The notional traveling flight path of the Martian sailplane is shown in Fig. 10. The craft holds the flight direction perpendicular to the wind direction in the atmospheric boundary layer or shear flows. Preforming cycles of dynamic soaring where energy gained during climb and descent, with the former being most effective (Fig. 10, 11). During this maneuver, the average flight altitude can be maintained or gained providing hours and days of flight by harvesting the energy from the atmosphere. Other flight paths, such as area scanning, loitering, and downwind advancing can be also realized by using dynamic soaring techniques (Fig. 11).

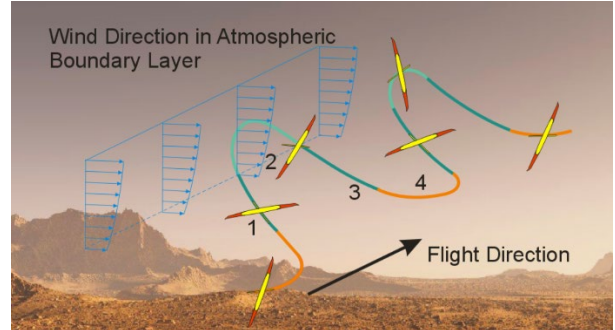

Fig. 10. Travel flight path (1 – windward climb, 2 – high-altitude turn, 3 – leeward descent, 4 – low altitude turn).

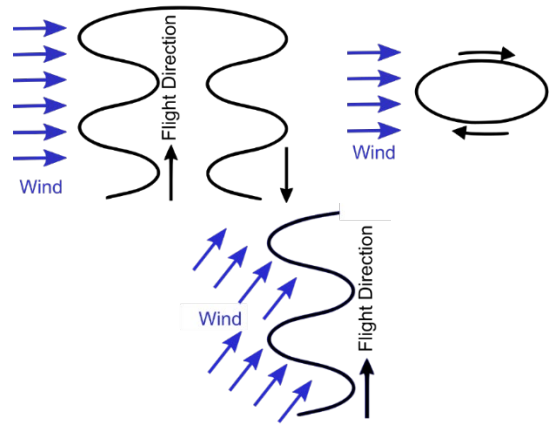

Fig. 11. Scanning (left), loitering (right), and downwind (bottom) flights.

The maximum energy harvesting climb ends when airspeed speed $V_a$ is close to the lower boundary $V_{a\,stall} = 51.4$ m/s, or if the wind gradient reaches below an operational limit set by aircraft performance, such that total energy decreases, $dE/dt < 0$. The sailplane initiates a downwind turn and descends to reach a the state vector projection $\vec{Y}_p = [z, V_a, \gamma, \psi]^T$ approximately equal to that of the initial point: $\vec{Y}_{pi} \approx \vec{Y}_{pe}$. The dynamic soaring cycle is referred to as successful if the total energy of the sailplane increases during the cycle, i.e.:

$\frac{E - E_{initial}}{E_{initial}} > 0$

Successful cycles are realized with control parameters $C_L$ and $\varphi$, from which the optimal trajectory can be found. This representation formulates the optimization problem, in which the objective is to maximize the relative kinetic energy rate along the flight path of the sailplane:

$$\max_{C_L,\varphi}\left\{\frac{d}{dt}(0.5mV_a^2): \forall t\right\} \quad (11)$$

subjected to boundary constraints: $V_{a\,target} = 55\frac{m}{s} \leq V_a \leq V_{a\,design\,limit}$, $-90^{o+} < \gamma < 90^{o-}$. Solution to the optimization problem gives an optimal flight trajectory of the sailplane executing the described dynamic soaring cycle.

Results from an MRAMS simulation campaign was used as the Mars boundary layer model, with a piecewise cubic interpolation of the discretized results expressing the wind vector as a function of $(x,y,z)$ and its spatial derivatives for the MRAMS result space.

Fig.12 shows such wind data spanning a day over the Jezero crater site. At this location high gradients exist in the atmospheric boundary layer below 1km and are sustained during the night with higher values in the late evening, the selected local times for dynamic soaring simulations are [18.25, 19.00, 20.00].

Segmented two-dimensional studies of a dynamic soaring cycle, and an uncontrolled spiral ($\gamma_0 = 30^o$, $v_0 = 55\,m/s$, $C_L = 0.45$, $\phi = 30^o$) (Fig. 13. top) show that the observed parabolic velocity distribution in the atmospheric boundary layer can yield increase in total energy, especial during the high altitude turn where an exchange in magnitude between the inertial and airspeed is observed (Fig. 12. bottom).

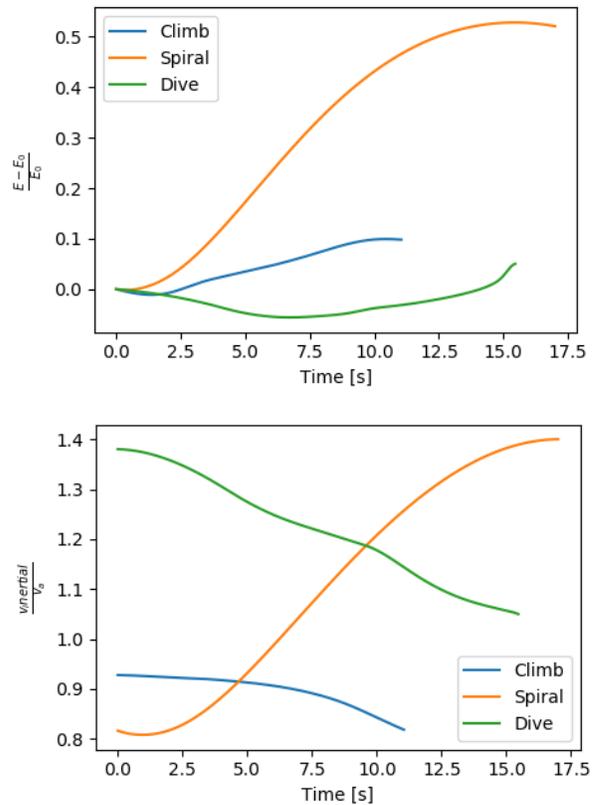

Fig. 13. Energy ratio (top), and inertial over aerodynamic velocities (bottom), from an itemized dynamic soaring cycle, wind at time 20.00 normalized to unidirectional

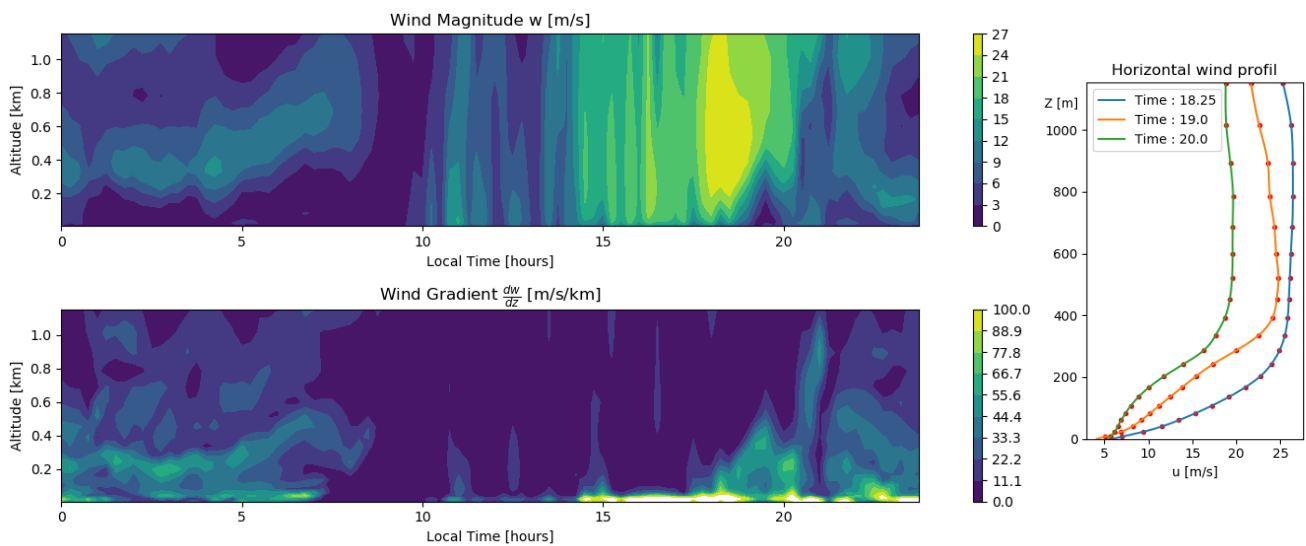

Fig. 12. Wind magnitude and gradient, one day of MRAMS results for Jezero crater (left), and wind magnitude profiles for selected dates (right).

Complete cycles present the overall feasibility of dynamic soaring on Mars, at the times selected for the most developed atmospheric boundary layer where there is evidence of successful cycles and sustained flight using this method (Fig. 14). For all three times the modeled cycles end with positive total energy ratios ([ 0.010, 0.050, 0.118, ]), for a maximum reached altitude of 454.6 m, 526.8 m, and 529.8 m, respectively, corresponding to the top areas of the atmospheric boundary layer velocity profiles observed in Fig. 12.

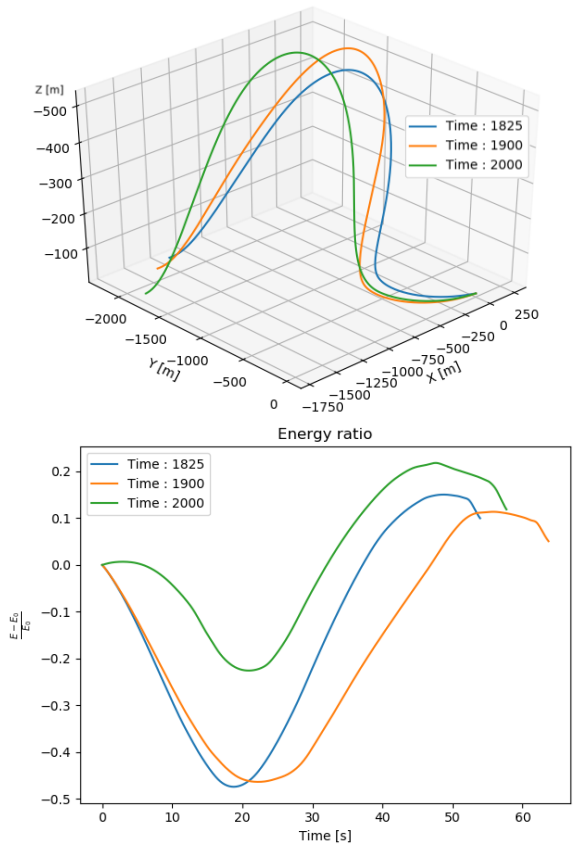

Fig. 14. Simulated dynamic soaring trajectory (top), sailplane total energy over time (bottom)

Energy is initially used to turn into the wind, this phase starts downwind and doesn't see a positive energy gradient, once the sailplane is heading upwind total energy increases in the climb, constituting the second phase. Once the top of the velocity profile is reached the sailplanes turns back downwind with a spiral turn, here energy is harvested as $V_a$ decreases and the inertial velocity $V_{inertial}$ increases. At the beginning of the dive downwind $V_a$ is close to stall, and increases during the decent, converting kinetic energy from the high inertial velocity into potential energy whilst still gaining from the wind. The cycle terminates with $z_{final} \sim z_{initial} = 10m$ and an increase of total energy through an increase in kinetic energy, and having advanced in both the crosswind and downwind direction

Dynamic soaring is then a viable solution for unpowered flight in proximity to the Martian terrain. The use of dynamic soaring to extending capabilities at targets associated with static soaring will be explored to further solutions for sustained flight at any time of the day with vertical or horizontal wind profiles.

9.  **Design of Sailplane and Deployment**

In our preliminary design, the Mars sailplane will be deployed during the descent phase, as the vehicle slows down to 80 m/s lateral velocity and achieves an altitude of 7 km (Fig. 2). The sailplane will undergo a 5-15 second deployment sequence at which it uses cold gas propulsion first to achieve a 1 km separation distance and achieve an altitude of 6 km to perform glide-cruise.

Two designs for quick deployment are being considered. One is a roll-up inflatable wing design. The second is an 'accordion' fold design of the inflatable wing. In the first design, the gas generation system from MER/Pathfinder heritage [TRL 9] will be packaged into a 6U form-factor CubeSat attached to another 6U contain all the remaining components of the aircraft (Fig. 15). The gas generation system will produce non-toxic $N_2$ and pressurize the wing within 3 seconds [32]. Telescopic booms developed by Oxford Space Systems [TRL 6] will expand to provide full structural support for the wing structure within about 10 seconds followed by disposal of gas generator and boom deployer. Inflatable structures have been demonstrated during EDL on Mars Pathfinder and Mars MER Rovers [33-35].

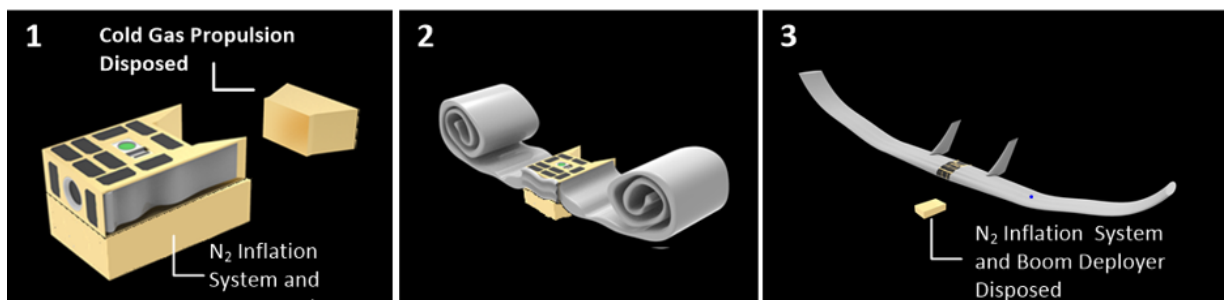

Fig. 15. Mars sailplane deployment steps. The cold-gas propulsion is used to achieve a separation distance (1) and then wing deployment begins (2,3).

The inflatable wing structure would contain a modular rib spar and a pair of extendable carbon-fiber Astrotube booms [36-38]. The Astrotube boom is shown to lift or hold a 300 g payload at its tip in Earth gravity. Each Astrotube boom would expand 2.85 m to provide support to the wing and expanded into place in 10 seconds.

A second design has some promising advantage over the roll-up design. This design can deploy fast and achieve a rigidized shape, as shown in Fig. 16 in the order of few seconds. The inflatable wing has a rigid support structure that starts off like an 'accordion' unfolds flat using torsional spring and rigidizes. As the inflatable system is proven to deploy in less than 3 seconds, we can expect a rigid wing be deployed within 3-5 seconds. The folded components swing open using torsional spring and lock into place. This is concurrently followed by the nitrogen filling in the inflatable.

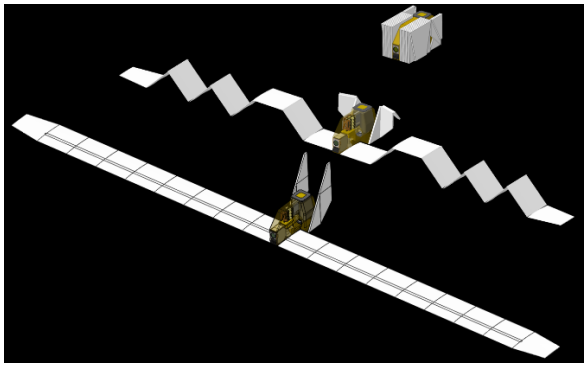

Fig. 16. Inflatable wing design using an accordion fold system.

A credible alternative to direct deployment of the sailplane during EDL is to first deploy a balloon [40] or blimp. These aircrafts would in turn carry the sailplane deployment package and deploy them at more leisurely pace in the order of hours. Balloon or blimp assisted deployment could significantly derisk the sailplane deployment sequence. As there is a large buoyant force available to keep the deploying Mars sailplane afloat in the atmosphere, there is less concerns for achieving lift, stability/control from the sailplane itself. Once the sailplane has deployed, it could be commanded to separate from the balloon/blimp perform dives and dynamic soaring maneuvers to perform science reconnaissance for scores of minutes to an hour and then come back and dock with the balloon or blimp. Hot air balloons have been proposed for Mars and require a device to heat the surrounding $CO_2$ atmosphere [40] (Fig. 17) [39]. The heating device could be powered using solar panels. The design could be used to generate enough power for staying aloft at night or gentle settle down/land at a location at night. A second option could be to deploy a blimp that can carry one or more sailplanes (Fig. 18) [39].

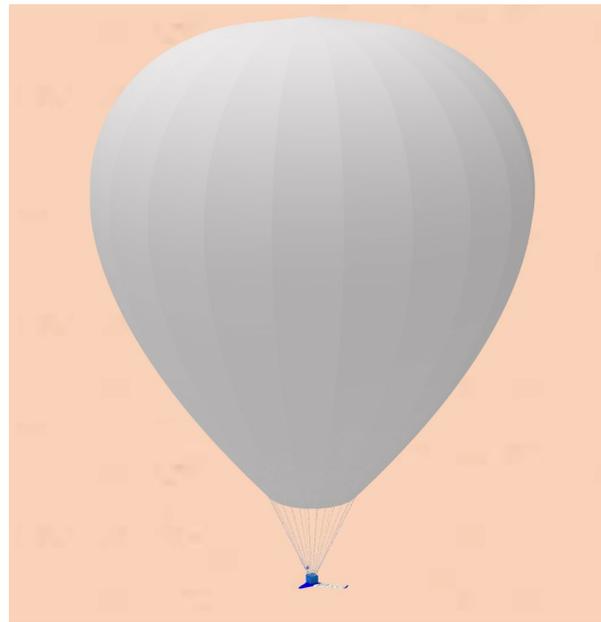

Fig. 17. Hot air balloon drawn to scale carrying a Mars Sailplane.

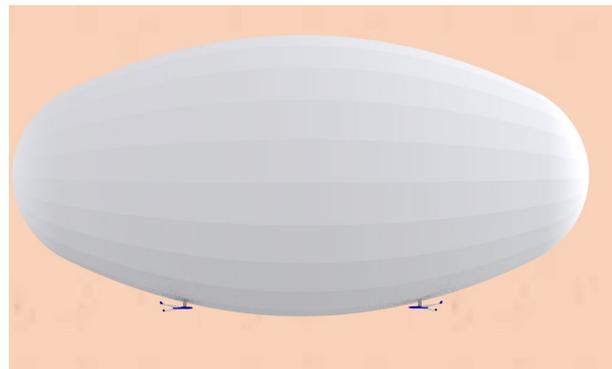

Fig. 18. Pair of Mars Sailplanes drawn to scale mounted to a powered blimp.

One or more sailplanes mounted to a powered blimp could have the advantages of the hot-air balloon discussed above in addition to the direction and altitude control of a blimp. Such a system has redundancies that could make long duration voyages lasting months to years credible. It can enable transport from the landing region (primarily located near the equator) to regions of interest such as Valles Marineris, Melas Chasma, Cydonia, Olympus Mons and Southern Highlands. The redundancies come from the use of two sail planes, where each one separates, uses dynamic soaring to explore surface targets of interest and then returns and docks with the hovering blimp. Even if one sailplane is lost, the mission continues. If there is any difficulty in reaching the blimp, the blimp in turn could also move towards the sailplane to enable a rendezvous maneuver. Furthermore, these rendezvous attempts can be carried out multiple times to ensure success. It should be noted that the blimp alone couldn't achieve what is possible with the blimp-sailplane hybrid system. The sailplane is smaller and has more maneuverability to do dives and get around edges of cliffs, canyons and deep valleys that would not be possible with the blimp alone.

### 10. Conclusions

The continued work on deployable sailplanes for the exploration of Mars has expanded the range of possible methods available to sustain perpetual flight through the harvesting of energy available in atmospheric flows near the surface. A preliminary design of a blended-wing-body sailplane with the symmetrical airfoil S9033 was modeled and proven capable of flying diving trajectories along a portion of the Valles Marineris, with results for optimal dive distances and velocities to then exploit slope or thermal flows efficiently. The formulated model was then applied to the dynamic soaring optimization problem, with boundary layer wind profiles from atmospheric flow simulations. The gain in total energy observed when utilizing local evening winds showed that this soaring method inspired by nature on earth is also valid for Mars, and can extend capabilities of atmospheric exploration missions in both time and space. Finally, various methods of deployment to Mars are presented including quick deployment during EDL and deployment using hot-air balloons and powered blimps. Using powered blimps it is possible to significantly derisk deployment challenges and transport/carry the sailplanes long distances to targeted regions, while exploit maneuverability of the sailplane(s) that would otherwise not be possible with the blimp alone to access steep canyons, cliffs and craters walls.